\documentclass[preprint,showpacs,preprintnumbers,amsmath,amssymb,superscriptaddress,prl]{revtex4}

\usepackage{graphicx}

\begin{document}

\title{Using synchronization to improve earthquake forecasting\\
in a cellular automaton model}

\author{\'Alvaro Gonz\'alez}
 \email{Alvaro.Gonzalez@unizar.es}
\affiliation{ Departamento de Ciencias de la Tierra. Universidad
de Zaragoza. Pedro Cerbuna, 12, 50009 Zaragoza, Spain.
}%

\author{Miguel V\'azquez-Prada}%
\affiliation{ Departamento de F\'isica Te\'orica and BIFI.
Universidad de Zaragoza. Pedro Cerbuna, 12, 50009 Zaragoza, Spain.
}%

\author{Javier B. G\'omez}
\affiliation{ Departamento de Ciencias de la Tierra. Universidad
de Zaragoza. Pedro Cerbuna, 12, 50009 Zaragoza, Spain.
}%

\author{Amalio F. Pacheco}
 \email{Amalio@unizar.es}
\affiliation{ Departamento de F\'isica Te\'orica and BIFI.
Universidad de Zaragoza. Pedro Cerbuna, 12, 50009 Zaragoza, Spain.
}%

\date{March 16, 2003}

\begin{abstract}
A new forecasting strategy for stochastic systems is introduced.
It is inspired by the concept of anticipated synchronization
between pairs of chaotic oscillators, recently developed in the
area of Dynamical Systems, and by the earthquake forecasting
algorithms in which different pattern recognition functions are
used for identifying seismic premonitory phenomena. In the new
strategy, copies (clones) of the original system (the master) are
defined, and they are driven using rules that tend to synchronize
them with the master dynamics. The observation of definite
patterns in the state of the clones is the signal for connecting
an alarm in the original system that efficiently marks the
impending occurrence of a catastrophic event. The power of this
method is quantitatively illustrated by forecasting the
occurrence of characteristic earthquakes in the so-called
Minimalist Model.\\
\end{abstract}


\maketitle

The last few decades have witnessed the irruption in geophysics
literature of many new concepts coming from modern statistical
physics, such as dynamical systems, chaos, fractals and self
organized criticality, among others \cite{TURCOTTE}. This has
been due to the genuine non-linearity of many geophysical
phenomena, and in some cases, a specific geophysical phenomenon
has been the origin of these nowadays widely known physical
concepts. In this sense, think for instance of the dynamical
meteorology equations in reference to chaos \cite{LORENTZ} or in
the Gutenberg-Richter distribution in seismology in reference to
self-organized criticality \cite{BAK}. In the study of complex
phenomena related to natural hazards (earthquakes, forest fires
and landslides), the so-called cellular automaton models have
resulted particularly useful \cite{MALAMUD}. On the other hand,
in spite of the unquestionable progress achieved by some groups,
one of the most important, everlasting, and difficult challenges
of geophysics is that of the prediction of large earthquakes
\cite{KEILISB,KOSSOBOKOV}.

We have recently introduced a model \cite{MM01} which tries to
simulate the dynamical routine in an individual seismic fault. It
is expressed by means of a one-parameter stochastic cellular
automaton which, for its simplicity, is called the Minimalist
Model (MM). The feasibility of forecasting the occurrence of the
largest earthquakes in the MM was assessed, in a first instance,
in Ref. \onlinecite{MM02}, where our results were certainly
modest. In this letter, we return to this endeavor, but now
equipped with a new, more powerful, method. It is inspired, in
part, by recent ideas on anticipated synchronization  between
chaotic oscillators, which constitutes a fascinating new topic in
the area of dynamical systems \cite{VOSS}. There, the oscillator
whose behavior one wishes to forecast is called \textit{the
master}, and the oscillator whose signal is equal to and precedes
that of the master is called \textit{the slave}. Its equivalent
here will be named \textit{the clone}. In spite of the obvious
differences between a chaotic oscillator and the discrete MM,
here we borrow several concepts from the former, which result in
being notably useful for our purposes.

\begin{figure}
\includegraphics{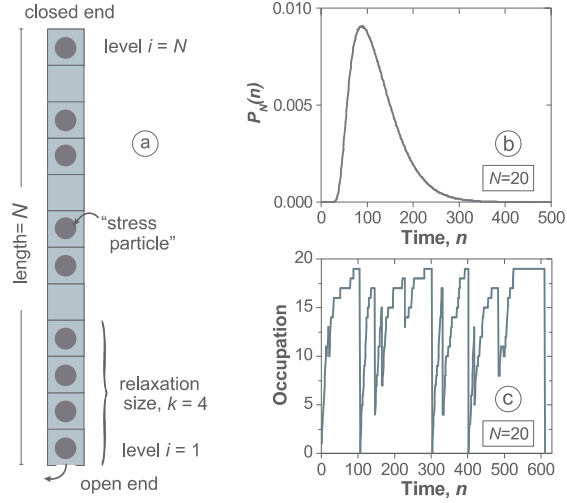}
\caption{\label{fig:1} (Color online) \textbf{(a)} Sketch of the
Minimalist Model (MM). \textbf{(b)} Probability distribution
function, $P_N(n)$, of the duration of the seismic cycle in a MM
of size $N=20$. \textbf{(c)} Evolution of the state of occupancy
during four seismic cycles in a MM of size $N=20$.}
\end{figure}

The MM \cite{MM01} was devised in a spirit akin to the sandpile
model of self-organized criticality \cite{BAK}: physics is not
apparent, but the model is able to grasp the basics of the
dynamical routine of an individual seismic fault. This model is
explained as follows (Fig. 1a). Consider a one dimensional
vertical array of length $N$. Its levels will be labeled by an
integer index $i$ varying from 1 to $N$. This system performs two
functions: it is loaded by receiving individual stress particles
in the various positions of the array; and unloaded by emitting
groups of particles through the first level, $i=1$, which are
called relaxations or earthquakes. These
two functions proceed according to the following four rules: \\
\textbf{1.} The incoming particles arrive at the system at a
constant time rate. Thus, the time interval between each two
successive particles is the basic time unit in
the evolution of the system.\\
\ \textbf{2.}  All the positions in the array, from $i=1$ to
$i=N$, have the same probability of receiving a new particle.
When a position receives a particle we say that it is occupied.
\\ \textbf{3.} If a new particle comes to a level which is
already occupied, this particle disappears from the system, so
its stress is dissipated. Thus, a given position, $i$, can only be
either empty, when no particle has come to it, or occupied when
one or more particles have come to it.
\\ \textbf{4.}  When a particle goes to
the level $i=1$, a relaxation event occurs. Then, if all the
successive levels from $i=1$ up to $i=k$ are occupied, and the
position $k+1$ is empty, the effect of the relaxation is to
unload all the levels from $i=1$ up to $i=k$, so the size of this
relaxation is $k$, $1 \le k \le N$. The remaining levels $i>k$
maintain unaltered their occupancy.  The prominent role given to
the level $i=1$ is analogous to that of a fault asperity (a
particularly strong element in the system which actually controls
its relaxation \cite{ASPERITY}).\\
Thus, this model has only one parameter: $N$, the size of the
array. The state of the system is given by stating which of the
$(i>1)$ $N-1$ levels are occupied. The earthquakes of maximum
size, $k=N$, are called characteristic, in analogy with the name
of the largest earthquakes an individual fault may produce
\cite{CHARACTERISTIC}. We will consider them as the target events
to forecast in the MM.

Earthquake forecasting strategies frequently consist in
connecting an alarm wisely before, and close to, the occurrence
of the target events, usually the biggest earthquakes
\cite{KEILISB,KOSSOBOKOV}. If an event occurs when the alarm is
off, this is a failure to predict. Conversely, if the event
occurs when the alarm is on, it is a successful prediction. Thus,
the \textit{fraction of error}, $fe$, is the ratio between the
number of failures to predict and the total number of target
events; and the \textit{fraction of alarm}, $fa$, is the ratio
between the time that the alarm was on and the total time of
observation. In order to evaluate quantitatively the forecasting
strategies that will be commented on later, we will use a loss
function, $L$, in the form: $L=fe+fa$. Other loss functions can
be chosen (see Ref. \onlinecite{KEILISB}, chapter 5), but this
will be used here for simplicity (as in Ref. \onlinecite{MM02}).
Our purpose is obviously to obtain a value of $L$ as low as
possible.

A seismic cycle of the MM lasts the time elapsed between two
consecutive characteristic earthquakes. Its duration is a
stochastic variable, with a probability distribution function
denoted by $P_N(n)$. In Fig. 1b, this distribution is plotted for
$N=20$. The knowledge of this distribution allows the setting of
a first strategy in forecasting (introduced in Ref.
\onlinecite{MM02}). It simply consists in connecting the alarm in
each cycle at a constant value of $n$ time steps after each
characteristic earthquake, and identifying the value of $n$ that
renders the lowest value of $L$. In Fig. 2, we have collected, for
comparison, the results of the various strategies in terms of
$N$. The curve labeled as ``Reference'' is the result of this
strategy, and will be considered as a baseline to assess the
actual merits of any other forecasting method \cite{NEWMANTUR}.

\begin{figure}
\includegraphics{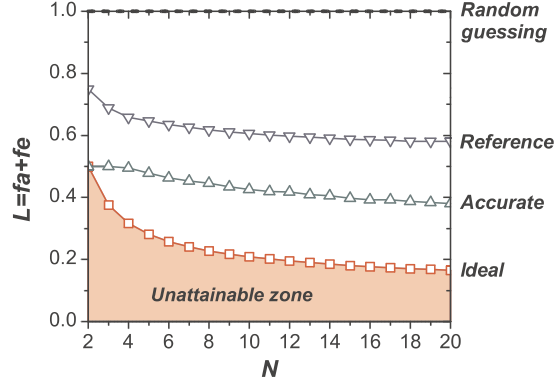}
\caption{\label{fig:2} (Color online) Loss function ($L$) obtained
with the three forecasting methods used in this letter, for
different system sizes ($2 \le N \le 20$). A random guessing
strategy would render $L=1$ for any $N$. The shadowed zone is
unattainable for any forecasting strategy used in the MM, and the
strategy that marks the lower limit of the reachable zone is
called ``Ideal''. The ``Reference''  strategy was presented in
\cite{MM02}, and the ``Accurate'' strategy is based on the
synchronization of copies (clones) of the system with the real
one (the master).}
\end{figure}

Now, it is important to realize that every cycle in the MM is
composed of two independent and consecutive stages (Fig. 1c). The
first one, that will be called the stage of \textit{loading},
starts just after the occurrence of a characteristic earthquake.
During this stage, the total number of occupied levels in the
system grows, but not in a monotonic way, because the particles
may be assigned to already occupied levels, and also because of
the occurrence of small relaxations. When the load accumulated in
the array reaches the maximum value of $N-1$ (when all the levels
but the first one are occupied), this first stage ends and the
second stage, that will be called the \textit{hitting} stage,
starts. In this second stage, the system resides in the state of
maximum occupancy until a particle hits the first level. Then, a
characteristic event occurs, all the load in the system is lost,
and a new cycle begins.

Both the time spent by the system in the loading stage, $x$, and
in the hitting stage, $y$, are statistically distributed. The
distribution of $y$, denoted by $P_2(y)$, is geometric. Its
density is $P_2(y)=(1/N)(1-1/N)^{y-1}$, and its mean is $<y>=N$.
As the variables $x$ and $y$ are independent, $<n>=<x>+<y>$. That
is, the mean length of the cycles $<n>$ is the sum of the mean
lengths of the two stages. Now, we deduce the best result one
could obtain in the forecasting of the characteristic events in
this model. It is clear that the best $L$ would be obtained
\textit{only if} we knew, in all the cycles, the instant at which
the system concludes the stage of loading. But this would imply
knowing the state of occupation of the system (\textit{i.e.}, the
state of stress at every point in the fault), which in real
seismic faults is obviously not possible. Here, however, and only
for the purpose of establishing this ideal efficiency, we will
assume that we know this information. Thus, we can explore the
result of $L$ if we connect the alarm at a fixed value $y=y_0$
within the second stage of the cycles. One easily deduces that
the minimum value of $L$ is obtained for $y_0=0$, \textit{i.e.}
just after the end of the first stage, which indicates that this
is a no-error forecasting strategy. The minimum value of L is
$L_{min}=N/<n>$, and constitutes a rigorous lower bound for the
accuracy of any forecasting strategy in the MM. The curve drawn
in Fig. 2 and labeled as ``Ideal'', represents this limit. For
this reason, the area below that curve has been darkened to mark
its inaccessibility.

Adopting the terminology used with chaotic oscillators, the
minimalist system whose characteristic events we want to forecast
will be called the master ($M$). It is important to realize that
$M$, by definition, is not manipulable at all. It is also
intrinsically opaque, in the sense that at any moment in the
cycle we know neither the value of the load in the system nor the
actual distribution of stress particles in the array. But, thanks
to the occurrence of relaxations in $M$, it is possible to deduce
some things about this distribution: just after the occurrence of
an earthquake of size $k$, at least the $k+1$ first levels of $M$
are empty.

Now, we define copies of $M$ which will be called $C$ (for
clones). In contrast to $M$, the $C$ systems will be transparent
and able to be manipulated; their only purpose is to help us to a
better forecasting of $M$. Thus, at any moment, we will know how
the load is distributed inside the clones and we will be able to
modify it at will. During the course of a cycle, in principle, we
will turn on the alarm when a certain pattern is observed in $C$,
but it will be turned off if the occurrence of a small earthquake
in $M$ disproves the risk of an imminent characteristic
earthquake. That would be a false alarm. Thus, during one cycle
several alarm intervals may be defined which contribute to the
fraction of alarm. $M$ and $C$ are driven subsequently, at the
same rate, during the whole duration of a cycle. This means that
for each stress particle randomly assigned to $M$, another one is
randomly assigned to each $C$. All these assignments are
independent of each other. Besides, the observed external effects
caused on $M$ by the addition of a particle (\textit{i. e.} the
possible occurrence of earthquakes in $M$) are dealt with first.
The new method of forecasting is explained
in six rules. Suppose, for simplicity, that only one clone is used.\\
\textbf{1.} If the particle assignment induces a relaxation of
size $k<N$ in $M$, we impose that $C$ evacuates its $k+1$ lowest
levels instead of receiving its corresponding particle
assignment.\\
\textbf{2.} If the particle assignment induces no relaxation in
$M$, we only allow the incoming particle in $C$ to be dropped in
any of the upper $N-1$ levels, so that we preclude the occurrence
of a relaxation in $C$.\\
\textbf{3.}  When, after a particle assignment, $M$ has no
earthquake and $C$ reaches its maximum state of occupancy, an
alarm is turned on in $M$. The cycle continues, and we wait until
the occurrence of the next earthquake in $M$. If this is a
characteristic one, then this event is added to the general
forecasting balance as a success, and its contribution to the
alarm fraction is registered. And a new cycle begins. If it is
not a characteristic one, pass to the fourth rule.\\
\textbf{4.}  If, as in the third rule, the complete filling of $C$
triggered an alarm in $M$ but the next earthquake is of size
$k<N$, this is a false alarm. So, the time elapsed since the alarm
was turned on is registered as a contribution to the alarm
fraction, the alarm is turned off, rule 1 is applied to $C$,
and the same cycle continues.\\
\textbf{5.}  If a characteristic earthquake occurs in $M$ before
$C$ has reached its maximum occupancy, \textit{i. e.} the alarm
was off, then this outcome is taken as a failure to predict.\\
\textbf{6.} Finally, if, as it will be described later, we are
using several clones, and one or more of them is completely full,
the remaining incompletely filled clones continue to be charged,
following the previous rules.

Note that these rules are designed in such a way that the
dynamical evolution of each clone tends to be synchronized with
that of the master. The complete filling of $C$ is interpreted as
an indication that $M$ is probably full, and, consequently, that
the occurrence of the characteristic earthquake is impending.
However, statistically speaking, each $C$ will complete the
loading stage before $M$. This is because, in the first rule, the
load lost by $C$ is less than or equal to $k$, while $M$, for
sure, loses $k$ units. When dealing with several $(Q)$ clones, $1
\le Q < \infty$, the six above mentioned rules for forecasting
are applied to all the clones in an identical form. Each clone
evolves independently to the others, and completes its total
occupancy at a different time. We have numerically explored, for
fixed $N$ and $Q$, what is the number of fully loaded clones,
$q$, one should wait for before connecting the alarm in $M$ and
obtaining the best prediction. This is illustrated in Fig. 3a for
$N=20$. For a fixed $N$, the value of $L$ improves as $Q$ grows.
However, the decrease in $L$ reaches a saturation as shown in
Fig. 3b, where an exponential fit has been performed to identify
the limit for $Q \to \infty$. These limit values for different
system sizes have been drawn in Fig. 2 forming the curve labeled
as ``Accurate''.

\begin{figure}
\includegraphics{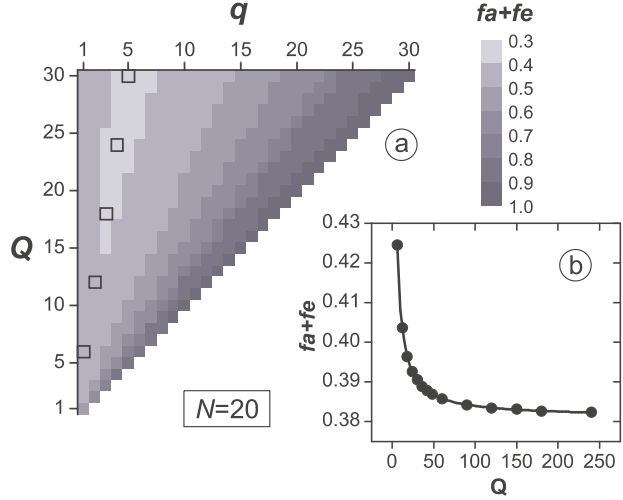}
\caption{\label{fig:3} (Color online) \textbf{(a)} Loss function
($L=fa+fe$) resulting from different combinations of $Q$, the
number of clones used, and $q$, the number of them awaited before
connecting the alarm in the master array, for a system of size
$N=20$. The marked cells indicate the combinations that mark the
lowest $L$ values per column $q$. \textbf{(b)} Minimum loss
function that may be obtained with a certain number $Q$ of clones
in a system of size $N=20$. The first five points correspond to
the marked cells of (a). The black dots result from Monte Carlo
simulations, and the solid line is an exponential fit, which has
the form $L=a\,\exp[b/(Q+c)]$, where $a$, $b$, and $c$ are
parameters. The extrapolation of this fit for $Q\to\infty$ renders
an asymptotic value of $L=a=0.381$, represented in the curve
``Accurate'' of Fig. 2 for $N=20$.}
\end{figure}

As far as we know, this is the first time that a forecasting
method for stochastic systems has been developed imitating the
anticipated synchronization of chaotic oscillators. Earthquake
forecasting algorithms frequently use several  pattern
recognition functions simultaneously, that, in real time, monitor
the seismicity in order to search for seismic precursors. When
several of them reach given thresholds, an alarm is turned on in
the zone under study \cite{KEILISB,KOSSOBOKOV}. This is the
inspiration for using several clones: they, in some way, monitor
the evolution of the master, and when a certain number of them
reach their maximum occupancy, an alarm is turned on in the
master. Our process of choosing the optimum $q$ is equivalent to
the learning phase of those algorithms \cite{KEILISB}. To leave a
quantitative flavor about the success of the clone-based
strategy, we will briefly compare its results with our previous
forecasting attempts on this model. In Ref. \onlinecite{MM02}, we
focused our attention on a system of size $N=20$.  Using the
strategy considered as baseline, the minimum loss function is
$L_{min}\mbox{\small{(reference)}}=0.578$. In Ref.
\onlinecite{MM02} we also tried a biparametric strategy
[$L_{min}\mbox{\small{(biparametric)}}=0.549$] and a triparametric
one [$L_{min}\mbox{\small{(triparametric)}}=0.528$]. So the
maximum improvement in \cite{MM02} was a modest 8.7\%, when
compared to the ``reference'' strategy. With the clone-based
method the minimum reachable value for $N=20$ (when $Q \to
\infty$) is $L_{min}\mbox{\small{(accurate)}}=0.381$ (Fig. 3b),
34.1\% better than the reference value, and halfway between the
latter and the ideal one (Fig. 2),
$L_{min}\mbox{\small{(ideal)}}=0.165$. Moreover, in the new method
the fraction of errors ($fe$) is systematically low: $fe=0$ for
$N=2$, and increases slowly with $N$, with $fe=0.095$ for $N=20$
and $Q \to \infty$. In any case, it is important to bear in mind
that the behaviour of a complex system cannot be predicted with
absolute precision. One can attempt to reduce the errors in the
forecasting, but not eliminate them completely \cite{KEILISB}.

\begin{acknowledgments}
Enlightening comments from Vladimir G. Kossobokov about the model
are warmly acknowledged. This research is funded by the project
BFM2002--01798 of the Spanish Ministry of Science. \'AG and MVP
hold the research grants AP2002--1347 (Spanish Ministry of
Education and Culture) and B037/2001 (Autonomous Government of
Arag\'on), respectively.
\end{acknowledgments}

%\bibliography{apssamp}

\end{document}